\documentclass[aip,showpacs,twocolumn,10pt]{revtex4}%
\usepackage{amsmath}
\usepackage{amsfonts}
\usepackage{amssymb}
\usepackage{graphics}
\usepackage{graphicx}
\providecommand{\U}[1]{\protect\rule{.1in}{.1in}}

\begin{document}
\title{Josephson directional amplifier for quantum measurement of superconducting circuits}
\author{Baleegh Abdo}
\altaffiliation{Current address: IBM T. J. Watson Research Center, Yorktown Heights, New York 10598, USA.}
\author{Katrina Sliwa}
\author{S. Shankar}
\author{Michael Hatridge}
\author{Luigi Frunzio}
\author{Robert Schoelkopf}
\author{Michel Devoret}
\email{michel.devoret@yale.edu}
\affiliation{Department of Applied Physics, Yale University, New Haven, CT 06520, USA.}
\date{\today }

\begin{abstract}
We have realized a microwave quantum-limited amplifier that is directional and can therefore function without the front circulator needed in many quantum measurements. The amplification takes place in only one direction between the input and output ports. Directionality is achieved by multi-pump parametric amplification combined with wave interference. We have verified the device noise performances by using it to readout a superconducting qubit and observed quantum jumps. With an improved version of this device, qubit and preamplifer could be integrated on the same chip.
\end{abstract}

\maketitle

\newpage

Quantum non-demolition (QND) measurements often require probing a quantum system with a signal containing only a few photons \cite{QuantumNoiseRev}. Measuring such a weak signal with high fidelity in the microwave domain involves a high-gain, low-noise chain of amplifiers. However, state-of-the-art amplifiers, such as those based on high electron mobility transistors (HEMT), are not quantum-limited \cite{Caves}; they add the noise equivalent of about $20$ photons at the signal frequency when referred back to the input. They can also have strong in-band and out-of-band back-action on the quantum system. In an attempt to minimize the noise added by the output chain, quantum-limited amplifiers such as the Josephson bifurcation amplifier (JBA) \cite{JBA,VijayJBAreview}, Josephson parametric amplifier (JPA) \cite{CastellanosNat,Yamamoto}, and Josephson parametric converter (JPC) \cite{Jamp,JRCAreview,JPCnature} have been recently developed and used as preamplifiers before the HEMT \cite{QuantumJumps,JPADicarloReset,SingleShotFluxQubit,QubitJPC}. Unfortunately, these quantum-limited devices amplify in reflection \cite{VijayJBAreview,JRCAreview,CastellanosAPL} and some of them have in addition strong reflected tones, e.g. reflected pump tone in the JBA case, which cause undesirable back-action on the quantum system. These devices also do not protect the measured system from back-action originating higher up in the amplification chain. Thus, non-reciprocal devices such as circulators and isolators are required in these measurements both to separate input from output, and also to protect the quantum system from unwanted back-action.

\begin{figure}
[b]
\begin{center}
\includegraphics[
width=\columnwidth 
]%
{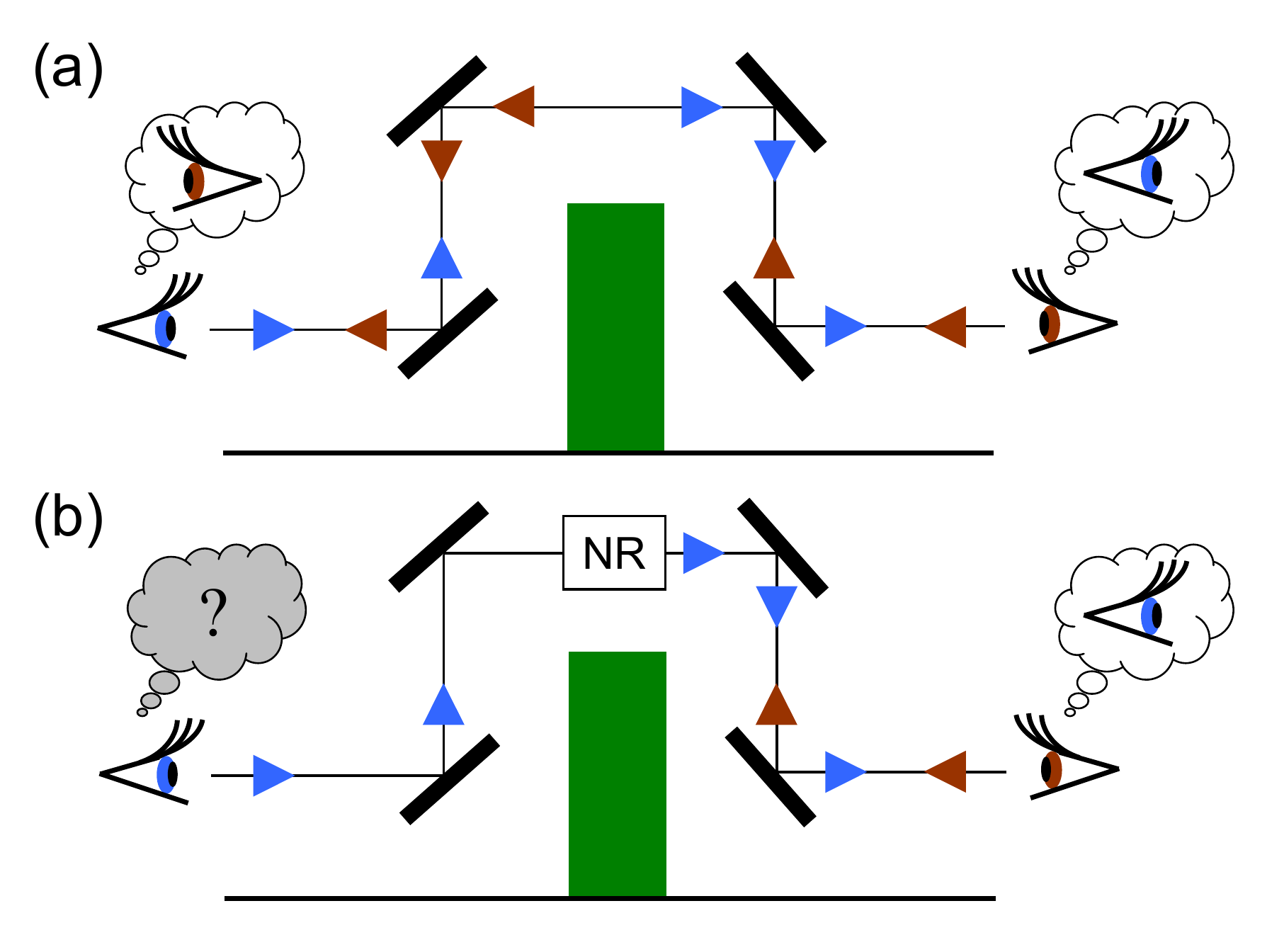}
\caption{(color online). (a) ``You see the eyes which see you": Fermat's second principle about the path of light being independent of its direction of propagation is one of the basic symmetries in wave physics known as reciprocity. (b) A medium or device which breaks this symmetry is called non-reciprocal (NR). The transmission coefficient through such medium/device depends on the direction of the waves. From Ref. \cite{archanaThesis}.
}
\label{Eyes}
\end{center}
\end{figure}

An illustration of the difference between reciprocal and non-reciprocal (NR) phenomena is shown in the cartoon in Fig. \ref{Eyes}. Panel (a) illustrates the reciprocity symmetry in wave physics, which is known in daily-life as ``you see the eyes which see you". In other words, a medium is reciprocal if its transmission coefficient for electromagnetic waves is invariant upon exchanging the source and the detector. Panel (b) on the other hand illustrates the less common phenomenon of non-reciprocity in which a certain medium or device breaks the reciprocity symmetry allowing only the detector to see the source and not the opposite. 

To achieve non-reciprocity, circulators and isolators exploit a magneto-optical effect known as Faraday rotation, which relies on ferrites and permanent magnets, in order to distinguish between polarized waves propagating in opposite directions \cite{Pozar}. Despite the reliability and good isolation of these components, utilizing them in low-noise quantum measurements of superconducting circuits has several drawbacks, (1) they are large in size, thus limiting the scalability of quantum systems, (2) their reliance on magnetic materials prevents integration on chip using present technology and makes them incompatible with superconducting circuits, and (3) they add noise to the processed signal as a result of their insertion loss. This loss is particularly critical when they precede the preamplifier. 

\begin{figure}
[t]
\begin{center}
\includegraphics[
width=\columnwidth 
]%
{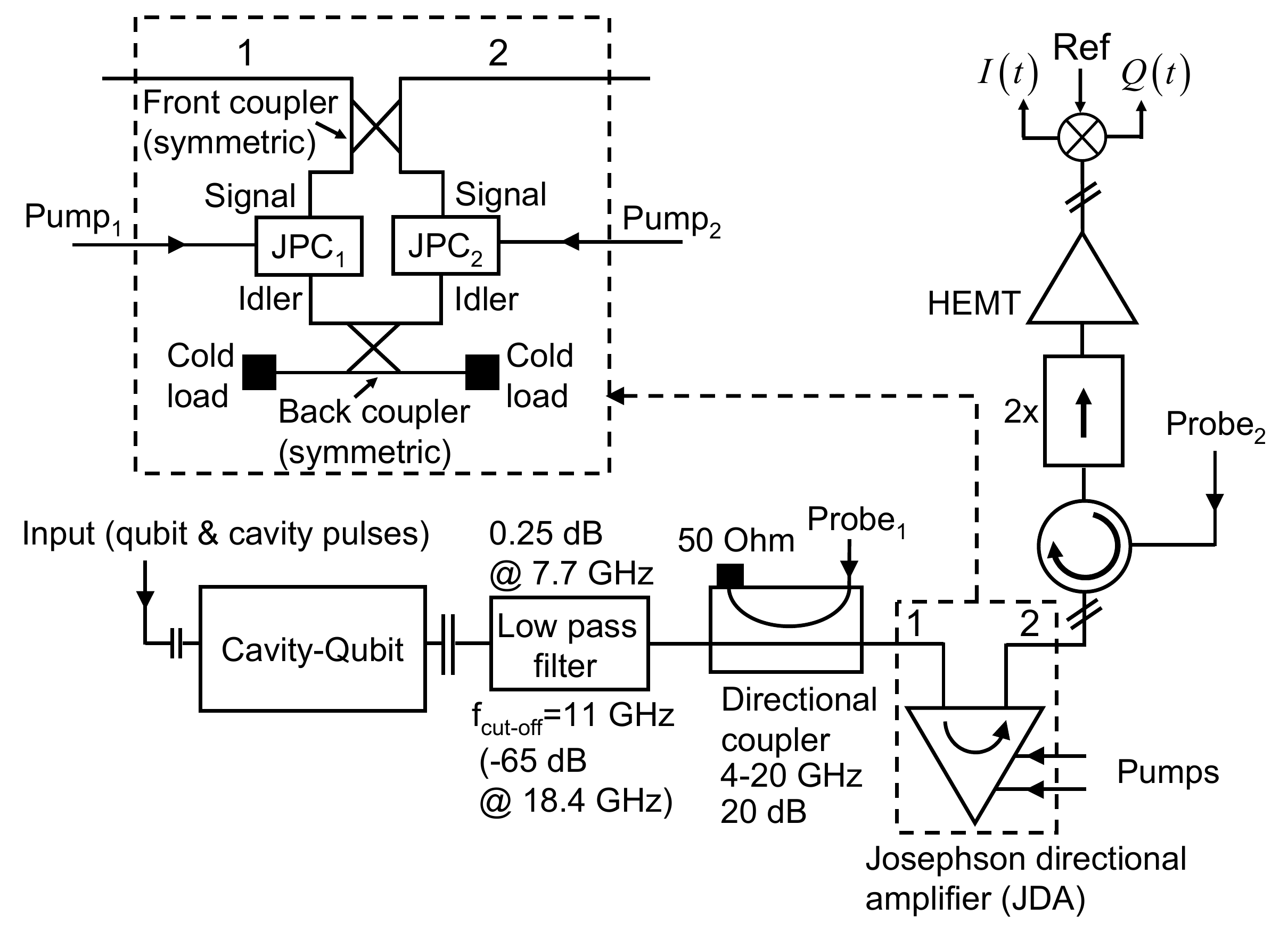}
\caption{Setup used in measuring the performances of the JDA whose schematic shown in inset. A superconducting cavity-qubit system is connected to the input of the JDA with no intermediate circulator or isolator stages. A low pass filter and a directional coupler are inserted in between the cavity-qubit system and the JDA. The filter protects the cavity-qubit system from high frequency pump photons that can possibly bounce off the JDA, while the directional coupler enables biasing the JDA at a desired working point using the input Probe$_{1}$. Similarly, the circulator connected at the output of the JDA enables measuring the reflection off port 2 of the JDA using the input Probe$_{2}$. The output signal is further amplified using a cryogenic HEMT amplifier and demodulated at room-temperature using an IQ mixer. 
}
\label{setup}
\end{center}
\end{figure}

Hence, the question which we are addressing in this letter is whether it is possible in principle to build a two-port directional amplifier based on the Josephson effect, which (1) works in the microwave domain (e.g. $4-12$ $\operatorname{GHz}$), (2) amplifies in one direction (i.e. non-reciprocal), (3) can amplify both quadratures of the microwave field (i.e. phase-preserving), (4) adds a minimum amount of noise required by quantum mechanics (i.e. quantum-limited), (5) matched to the input and output, (6) have as little in-band and out-of-band back-action on quantum systems as possible, and (7) have sufficient gain, bandwidth, frequency tunability, and dynamic range comparable to existing state-of-the-art parametric amplifiers \cite{JRCAreview,JBAMichael}.    

As we show in this work, the answer for this question is affirmative: yes. There are other implementations of directional amplifiers
but they have so far met difficulties in satisfying all the aforementioned criteria. For example, SQUID-based directional amplifiers, such as the microstrip SQUID amplifier (MSA) \cite{Spietz,MSAnegativeFB,Kinion} and the superconducting low-inductance undulatory galvanometer (SLUG) \cite{SLUGThy,SLUGExp}, dissipate energy on chip and have out-of-band back-action which still requires using circulators in quantum measurements \cite{MSAreadout}. There is also a different type of directional amplifier known as the traveling-wave parametric amplifier in which the nonlinearity of kinetic inductance of superconducting transmission lines \cite{CaltechAmp} or Josephson junctions \cite{TWPIrfan} is exploited in order to parametrically amplify weak propagating signals. To the best of our knowledge, the added noise and back-action of this type of amplifier has not been fully characterized in a quantum measurement yet. 

The Josephson directional amplifier (JDA) consists of two nominally identical JPCs coupled together through their signal and idler ports \cite{SM}. As shown previously, coupling two JPCs together can, under certain conditions, generate directional amplification for a certain applied pump phase difference between the two stages \cite{DircJPC}. However, despite the relatively good performance of the proof-of-principle device presented in Ref. \cite{DircJPC} with respect to directionality, bandwidth, gain, dynamic range, and added noise, it had three main shortcomings, (1) the reflection gain at the desired working point is suppressed only on one port of the device, (2) the coupling between the two stages, which determines the device directionality and performance, varies with the flux and pump frequency, thus finding a good working point can be complicated, and (3) the back-action of the device cannot be easily modeled or characterized.

\begin{figure}
[b]
\begin{center}
\includegraphics[
width=\columnwidth 
]%
{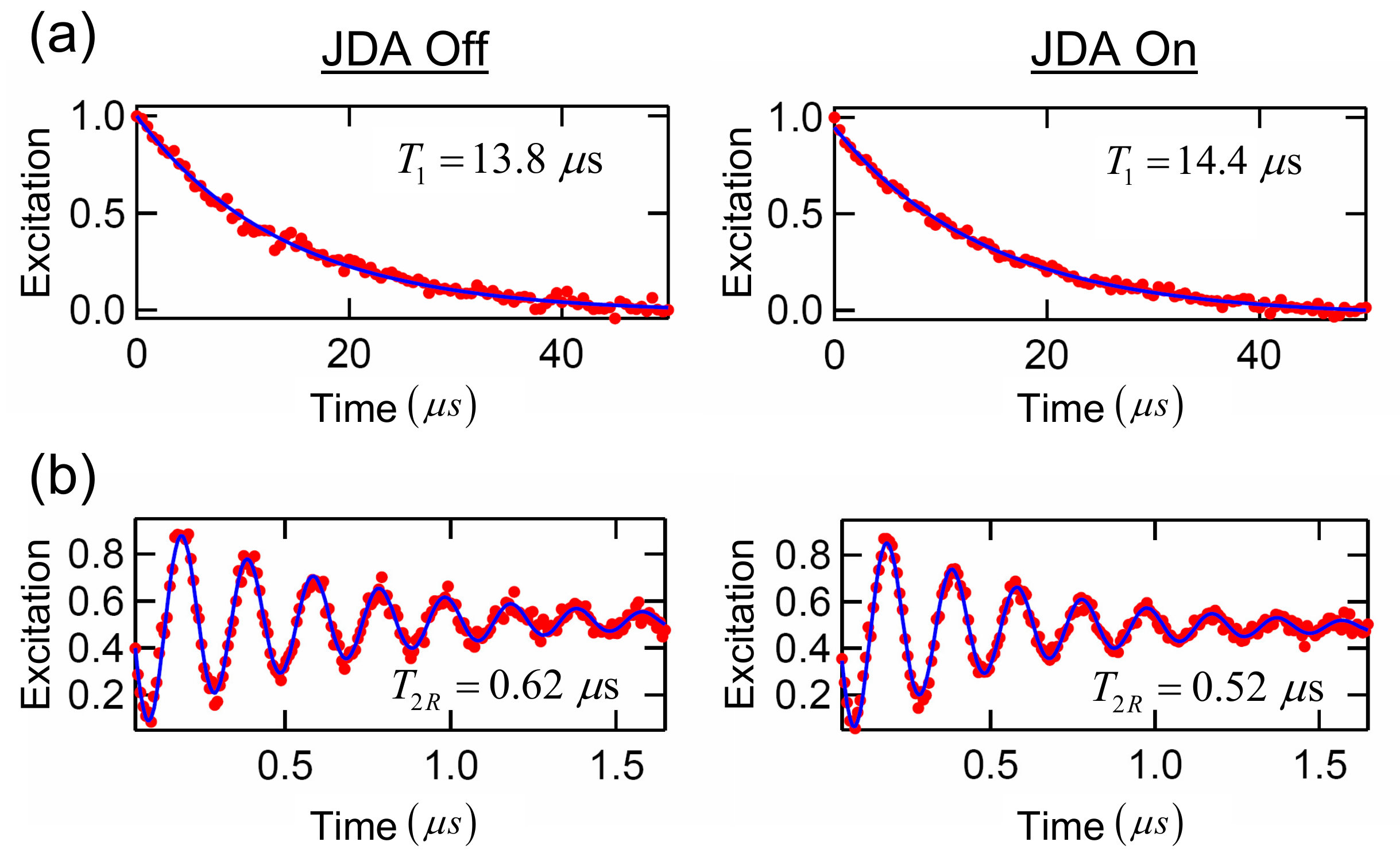}
\caption{(color online). A comparison between the decoherence times of the qubit measured with the JDA off (left column) versus on (right column). Plots (a) and (b) show $T_{1}$ and $T_{2}$ Ramsey measurement results respectively. The red filled circles are data points while the blue curves are fits. From the fits we get $T_{1}$ of $13.8$ $\operatorname{\mu s}$ and $14.4$ $\operatorname{\mu s}$, and $T_{2R}$ of $0.62$ $\operatorname{\mu s}$ and $0.52$ $\operatorname{\mu s}$ for JDA off versus on.   
}
\label{T1T2}
\end{center}
\end{figure}

\begin{figure*}
[t!]
\begin{center}
\includegraphics[
width=1.5\columnwidth 
]
{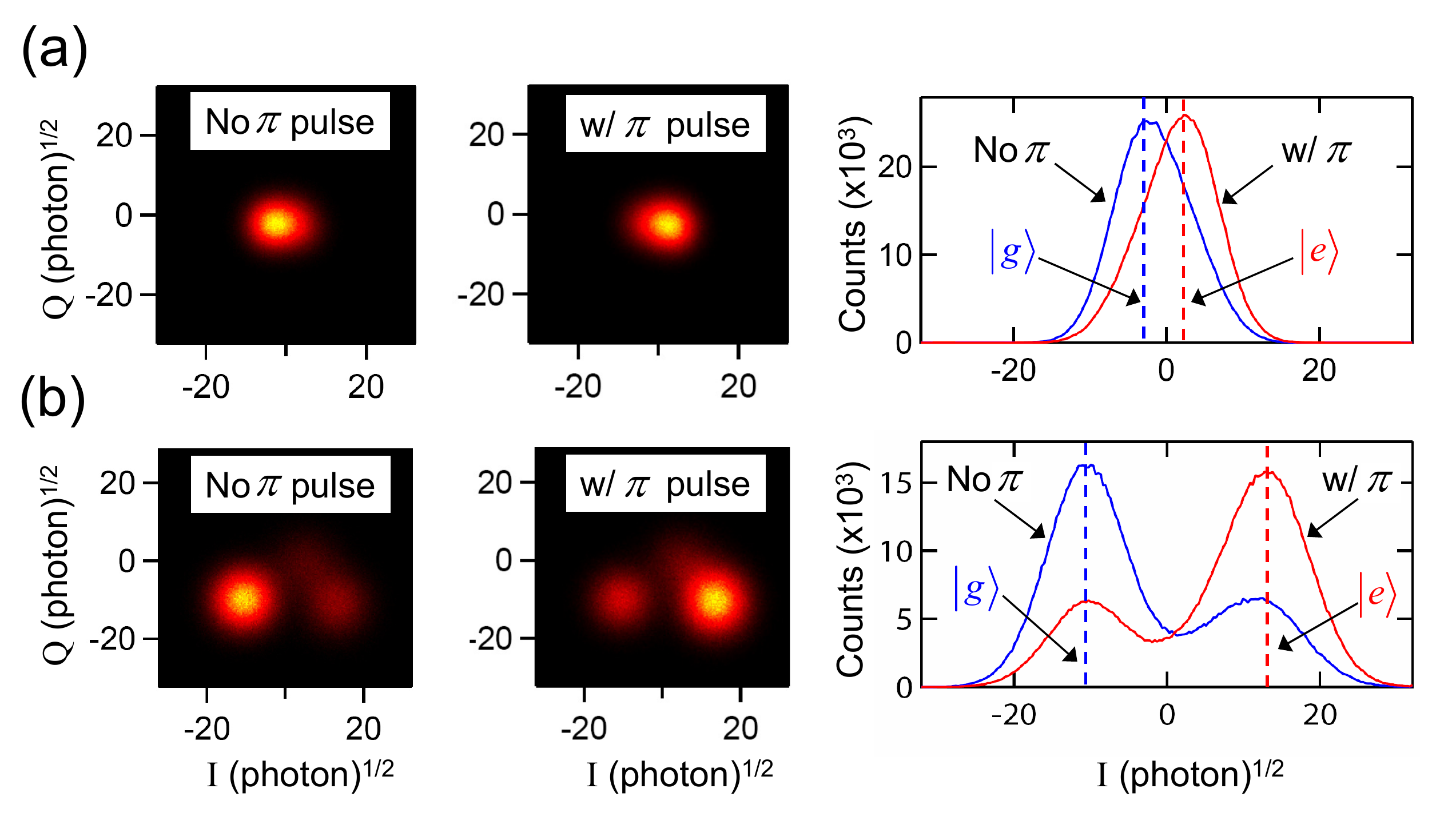}
\caption{(color online). Histograms of the qubit state measured after applying a $\pi$ pulse or no pulse to the qubit. Panels (a) and (b) show the histograms measured with the JDA off and on respectively. In the left and middle columns, we depict 2D histograms of the qubit state as a function of the $I$ and $Q$ quadratures of the microwave field, expressed in units of square root of average number of photons in the readout pulse at the output of the JDA. The right column depicts projections of the 2D histograms for the ground and excited states onto the $I$ quadrature.
}
\label{histograms}
\end{center}
\end{figure*}

In this work, we have developed a novel coupling scheme between the two JPCs which meets the requirements for quantum measurements, resolves the problems of the previous design, and leads to a calculable amplifier \cite{SM}. The basic idea of the JDA is to transform the non-reciprocal phase of the JPC in the transmission-gain mode, which is set by the phase of the pump drive, into a non-reciprocal amplitude response in the full device, by way of wave interference between different paths. These are formed by coupling two JPCs with the same characteristics together through their signal and idler ports. The signal ports are coupled through a symmetric 50/50 beam-splitter (i.e. 90 degree hybrid), while the idler ports are connected together in a feedback loop with certain attenuation. The beam-splitter on the signal ports serves as an interferometer, which splits and combines incoming and outgoing waves, and nulls reflections under certain conditions. The feedback loop on the idler side with the attenuation is essential in order obtain net gain in the device. As shown in the supplemental material (SM) directional amplification can be generated between ports 1 (input) and 2 (output) of the device, by operating the two JPCs in amplification mode with the same gain and applying a phase difference of $\pi/2$ between the coherent pump drives feeding the two stages.

\begin{figure}
[htb]
\begin{center}
\includegraphics[
width=\columnwidth 
]%
{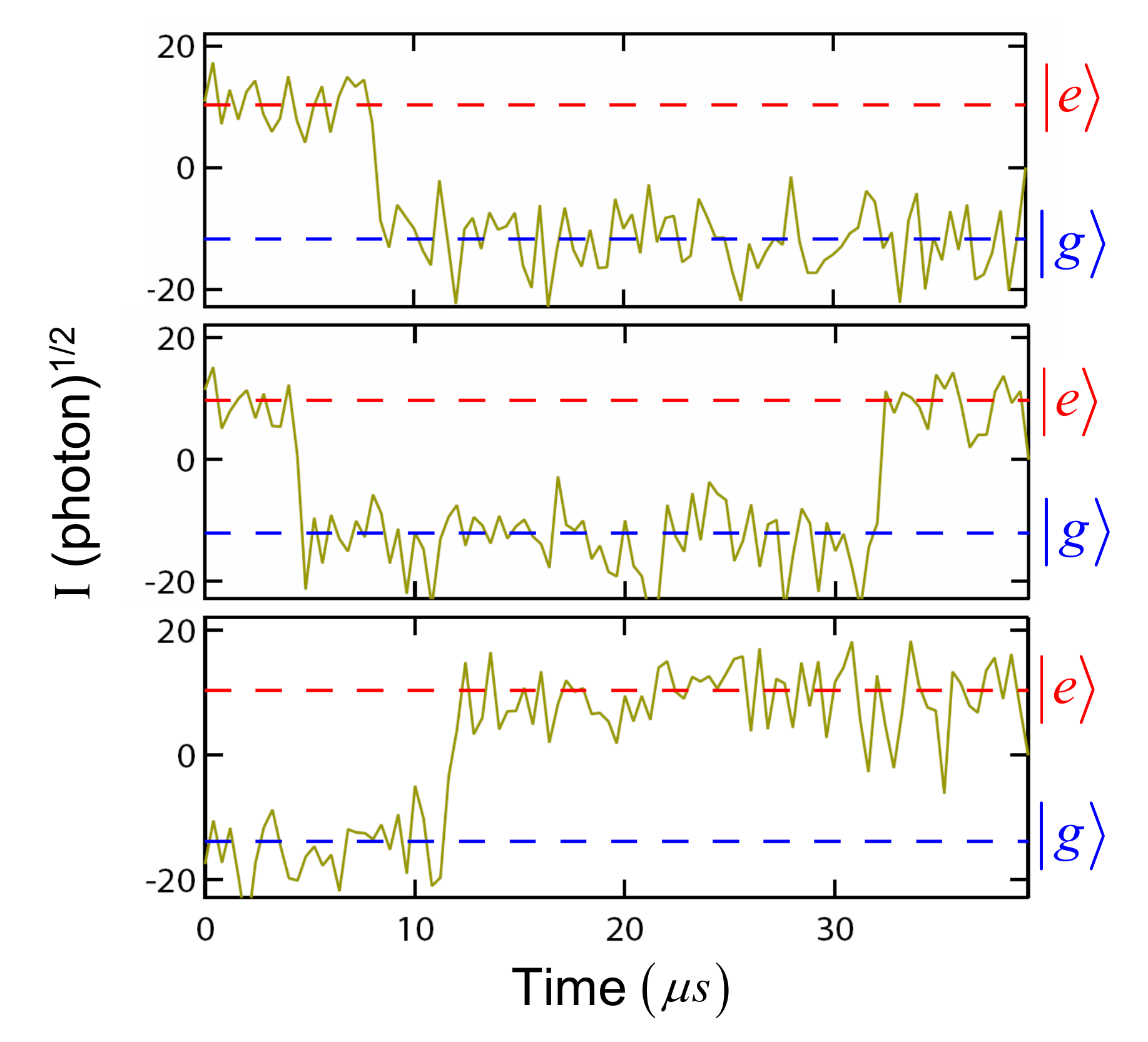}
\caption{(color online). Selected individual measurement records of the JDA-on output. The qubit state traces versus time exhibit quantum jumps between the ground and excited states of the qubit. The red and blue dashed lines which indicate the excited and ground signal levels of the qubit respectively, are guides for the eye.
}
\label{Jumps}
\end{center}
\end{figure}

To verify that the JDA meets the requirements for quantum measurements, we have coupled our JDA to a superconducting qubit-cavity system without any intermediate circulators, as shown in Fig. \ref{setup}, and used it in order to perform high-fidelity QND measurements of the qubit state (see Figs. \ref{T1T2} and  \ref{histograms}). Using this setup, we were also able to observe quantum jumps between the ground and excited states of the qubit (see Fig. \ref{Jumps}). 

The cavity-qubit system used in the experiment is an aluminum 3D cavity containing a single junction transmon \cite{Hanhee}. The resonance frequency of the cavity when the qubit is in the ground state and the cavity bandwidth are $f_{r,g}=7.7132$ $\operatorname{GHz}$ and  $\kappa/2\pi=3.4\operatorname{MHz}$ respectively. The qubit frequency is  $f_{eg}=4.35181$ $\operatorname{GHz}$ and the state-dependent shift of the resonator frequency is $\chi/2\pi=3.7 \operatorname{MHz}$. In the experiment, the readout tone frequency  $f_{d}=7.71135$ $\operatorname{GHz}$ was set to be at the center between the state dependent resonances $f_{r,g}$ and $f_{r,e}$, and its power was set to yield an average number of photons in the cavity of $2.9$ during the readout pulse of length $T_m=250$ n$\operatorname{s}$.  

To readout the qubit state, the center frequency of the JDA was tuned to $f_{d}$ (the corresponding center frequency on the idler is about $10.75$ $\operatorname{GHz}$). This was achieved by varying the flux threading the 8-Josephson junction ring modulators of the two JPC stages \cite{Roch,JRCAreview}. To operate the JPCs in amplification mode, the frequency of the pump drive feeding them, which is a coherent nonresonant tone, was set to $f_p=18.464$ $\operatorname{GHz}$, i.e. the sum of the center frequencies of the signal and idler resonators \cite{JRCAreview}. To improve stability, the pumps applied to both JPCs are generated from a split single generator, with one arm passing through a variable phase shifter and attenuator. The attenuator was inserted in order to compensate for the different amount of attenuation on the pump lines, while the phase shifter was used in order to maximize the directionality of the JDA in the forward direction ($|S_{21}|^2$) for a certain pump amplitude, and minimize at the same time the reflection amplitude on port 2 ($|S_{22}|^2$). Such in-situ tuning of the JDA working point was enabled by the addition of a directional coupler at the input of the JDA and a circulator at its output, as can be seen in Fig. \ref{setup}. In addition, a low-pass filter with a cut-off at $11$ $\operatorname{GHz}$ was added between the cavity and the JDA  mainly as a protection for the cavity-qubit system from pump photons that can bounce off the JDA due to the finite isolation ($20$ dB) between the $\Sigma$ and $\Delta$ ports of the 180 degree hybrids used for feeding the pumps to the JDA (see Fig. S6 in the SM). The working point of the JDA at which the data in Figs. \ref{T1T2}, \ref{histograms}, \ref{Jumps} was taken is shown in Fig. S7 in the SM. The forward gain of the device is set to $11.6 \pm1$ dB with dynamical bandwidth of $15$ $\operatorname{MHz}$ \cite{SM}. It is important to note that the qubit-cavity can be measured with the JDA off also. That is because the JDA has a unity transmission in that case \cite{SM}.

In Fig. \ref{T1T2}, we show the relaxation time $T_1$ and $T_{2R}$ of the qubit measured with the JDA off versus on. As can be seen in the figure, turning the JDA on improves the measurement signal to noise ratio (SNR) without any observed degradation in the relaxation time of the qubit or considerable effect on $T_{2R}$. The fact that the JDA does not alter $T_1$ provides an important piece of evidence that it can perform QND measurements. The relatively short $T_{2R}\simeq0.6$ $\operatorname{\mu s}$ in both cases can be attributed to the relatively high base temperature of the dilution fridge in the experiment $54$ m$\operatorname{K}$, as well as to the relatively short inverse residence rate of photons in the cavity $1/\kappa=47$ n$\operatorname{s}$. Knowing $\kappa$ and $T_{2R}$, we can calculate the maximum average number of thermal photons in the cavity required in order to account for the measured $T_{2R}$, which is given by $1/\kappa T_{2R}=0.08$. Similarly, by subtracting the ratio $1/\kappa T_{2R}$ for the JDA on versus off cases, we can set a bound of about $0.01$ added photons in the cavity due to the back-action of the JDA. Such back-action effect can be due to pump power of about -150 dBm that leaks back from the JDA and populate a higher resonance mode of the cavity or due to low-gain amplified quantum noise which is transmitted in the opposite direction ($S_{12}$) at resonance. 

To further verify that the JDA performs a QND operation, we have measured the qubit population in the ground ($p_{g}$) and excited ($p_{e}$) states with the JDA on versus off using the measurement method presented in Ref. \cite{ResetProtocolKurtis}. In both measurements, we obtained the same nominal result $p_{g}=0.8\pm0.02$ and $p_{e}=0.2\pm0.02$ which shows that the JDA does not affect the qubit population. Furthermore, from these measurements we extract a qubit temperature of $T=154\pm4$ m$\operatorname{K}$.  
 
In Fig. \ref{histograms} on the left, we depict 2D histograms of single shot measurements of the qubit state obtained upon applying a $\pi$ pulse or no pulse to the qubit. In panels (a) and (b) we plot the histograms measured with the JDA off and on respectively in both quadratures of the microwave field, i.e. $I$ and $Q$, expressed in units of square root of average number of photons in the readout pulse at the output of the JDA. As expected, without applying a pulse (left column) and following a $\pi$ pulse (middle column), the majority of the qubit population is in the ground and excited states respectively. The main differences between the histograms shown in panel (b) compared to (a) are that the histograms in (b) are broader, more separated, and a small population in the $f$ state is visible close to the origin while it is hidden in panel (a). In the right column, we plot projections of the histograms corresponding to the ground and excited states onto the $I$ quadrature axis. As can be seen in the bottom plot, with the JDA on, the Gaussian curves corresponding to the $g$ and $e$ states are sufficiently separated (about $3.7\sigma$) which allows us to perform single shot measurements of the qubit state. Moreover, from the standard deviations of the measured histograms for the JDA off ($\sigma_{\mathrm{Off}}$) and on ($\sigma_{\mathrm{On}}$) cases, we find that the number of noise photons added by the output chain (with the JDA off) and by the JDA at the signal frequency when referring back to the input are $n_{\mathrm{sys}}=17\pm1$ and $n_{\mathrm{JDA}}=0.9\pm0.4$ respectively \cite{SM}. The result for $n_{\mathrm{sys}}$ agrees well with our prior knowledge of the noise temperature of the output line which is $6.5\pm1$ $\operatorname{K}$. Also as predicted by theory \cite{SM}, $n_{\mathrm{JDA}}$ shows that the JDA indeed operates near the quantum-limit which is $1/2$ for phase-preserving amplification \cite{Caves}.   

In addition, we measured 2D histograms of the qubit state for different drive powers. This measurement allowed us to determine the onset of input power at which the JDA start to saturate as reflected from the histogram separation (which stops scaling as the square root of the input power). Using this method, we find that this power is about $7$ photons at the readout frequency per inverse cavity bandwidth. This figure of merit can be improved by enhancing the dynamic range of the JPC stages that form the JDA \cite{JRCAreview}. It is also important to point out that the JDA gain attained in this experiment is mainly limited by non-idealities in the cavity-qubit setup, such as using infra red filters on the pump lines which are not well matched to $50$ Ohm \cite{SM}. Fixing these problems should allow the JDA to operate at gains in excess of $20$ dB. 
 
In Fig. \ref{Jumps} we display three selected individual measurement records of the qubit state. The data are digitized with a sampling time of $20$ n$\operatorname{s}$ and smoothed with a boxcar filter with $400$ n$\operatorname{s}$ width, which corresponds to $8.5$ cavity lifetimes. As can be seen in the figure, the traces of the qubit state versus time exhibit quantum jumps between the $g$ and $e$ states. It also shows that with the JDA we can perform single shot measurements of the qubit state and track its state in real time, which are essential requirements for many quantum feedback applications \cite{RabiVijay,JPADicarloFB}.  

In conclusion, we have realized a novel Josephson directional amplifier suitable for quantum measurements and demonstrated its performances by reading out the quantum state of a superconducting qubit. Looking forward, an on-chip version of the directional amplifier presented in this work can be in principle implemented using standard microwave technology and fabrication processes. Moreover, the ability to perform high-fidelity, QND measurements without incorporating intermediate circulator stages between the qubit and the quantum-limited amplifier opens the door for realizing quantum systems with high measurement efficiency, in which qubits and preamplifiers are integrated on the same chip.
 
\begin{acknowledgments}
Discussions with Anirudh Narla are gratefully acknowledged. This
research was supported by the NSF under grants DMR-1006060 and DMR-0653377; ECCS-1068642, the
NSA through ARO Grant No. W911NF-09-1-0514, IARPA under ARO Contract No.
W911NF-09-1-0369.
\end{acknowledgments}

\end{document}


\title{Supplemental Material for ``Josephson directional amplifier for quantum measurement of superconducting circuits''}
\author{Baleegh Abdo}
\altaffiliation{Current address: IBM T. J. Watson Research Center, Yorktown Heights, New York 10598, USA.}
\author{Katrina Sliwa}
\author{S. Shankar}
\author{Michael Hatridge}
\author{Luigi Frunzio}
\author{Robert Schoelkopf}
\author{Michel Devoret}
\email{michel.devoret@yale.edu}
\affiliation{Department of Applied Physics, Yale University, New Haven, CT 06520, USA.}
\date{\today }

\maketitle

\section{General scattering matrix of the Josephson directional amplifier}
\begin{figure}
[h]
\begin{center}
\includegraphics[
width=0.55\textwidth
]%
{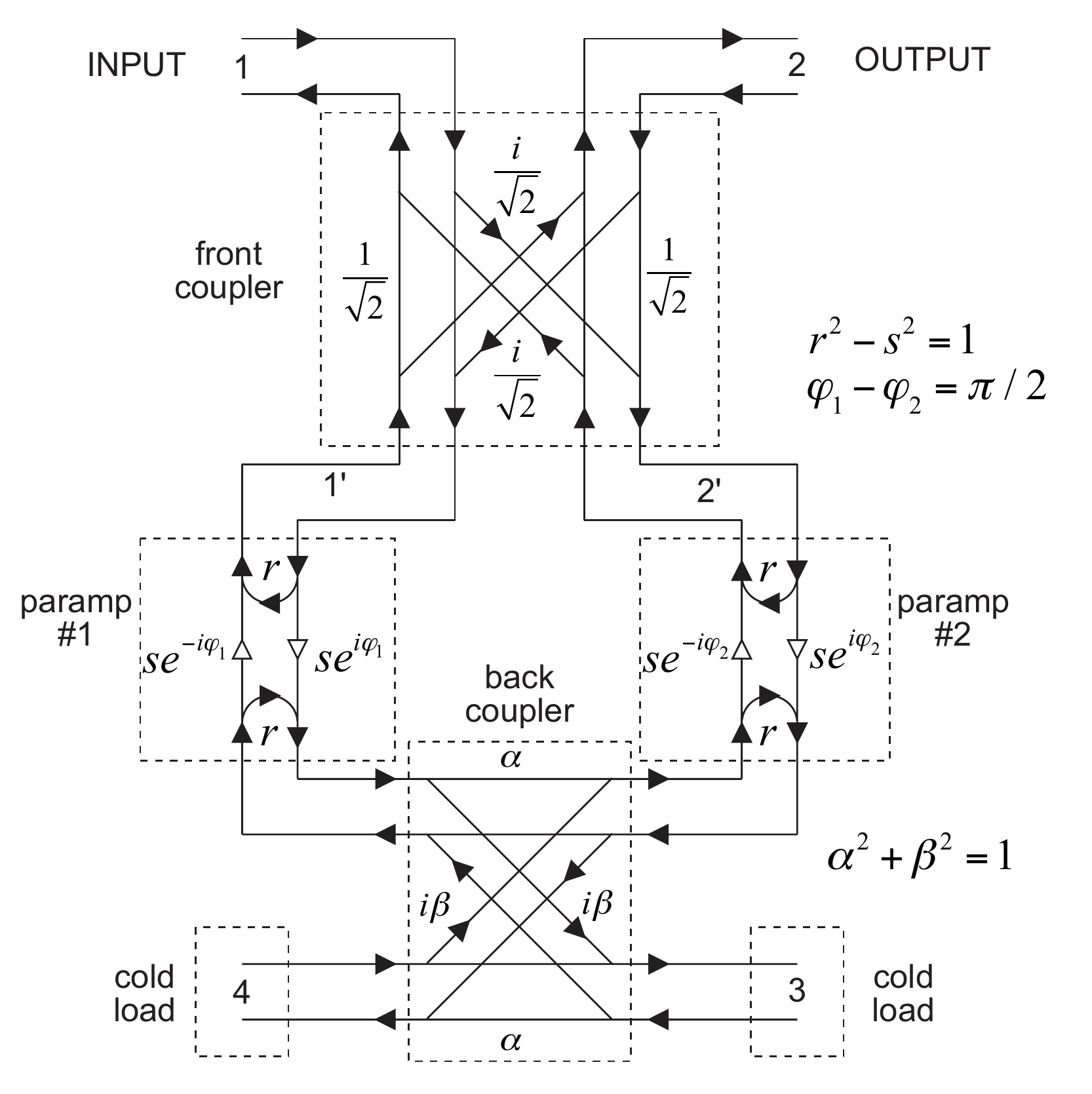}%
\caption{A signal flow graph of an ideal model of the JDA used in the experiment. The model comprises of two identical paramps coupled together through their signal and idler ports using two symmetrical hybrids referred to as front and back couplers respectively. The front coupler is taken with equal splitting ratio, while the back coupler has generic real coefficients $\alpha$ and $\beta$, which satisfy the condition $\alpha^{2}+\beta^{2}=1$. Ports 1 and 2 correspond to the input and output ports of the device, while ports 3 and 4 correspond to cold loads (dumps). One viable candidate for a paramp in this scheme is the Josephson parametric converter (JPC) \cite{Jamp,JRCAreview}. Directional amplification can be generated in the device by applying a phase gradient ($\varphi_{1}-\varphi_{2}$) between the pump tones feeding the two paramps. Maximum forward gain (from input to output) can be obtained for $\varphi_{1}-\varphi_{2}=\pi/2$.}%
\label{DAScheme}%
\end{center}
\end{figure}

The schematic of the Josephson directional amplifier (JDA) showing its internal wave propagation is depicted in Fig. \ref{DAScheme}. In order to obtain the scattering matrix of the device, we solve equations obtained by following signal flow. The JDA consists of two paramps (parametric amplifiers) coupled together via two symmetrical couplers, which we refer to as front and
back couplers. The front coupler is a $3$ dB coupler, which couples the input
and output ports of the JDA (ports $1$ and $2$) to the signal ports of the
paramps (ports $1^{\prime}$ and $2^{\prime}$). The back coupler having generic real coefficients $\alpha$ and $\beta$ (satisfying the relation $\alpha^{2}+\beta^{2}=1$), couples the idler ports of the paramps to cold loads (dumps) indicated in the figure as ports $3$ and $4$. The arrows in the signal graph indicate the propagation direction of the waves \cite{Pozar}. Their black or white
color indicates whether the wave frequency remains the same or undergoes
frequency conversion with conjugation. We also denote the reflection-gain and
trans-gain amplitudes of the paramps at resonance as $r$ and $s$ respectively,
where $r^{2}-s^{2}=1$. For simplicity, this model assumes balanced paramps,
having the same $r$, $s$ and the same characteristics. However, the coherent
pump tones feeding the paramps can have different phases, denoted as
$\varphi_{1}$ and $\varphi_{2}$ for paramp $1$ and $2$ respectively. In
addition, in order to ensure stability in the feedback loop of the JDA formed by the back
coupler, we bound the reflection-gain amplitude $r$ within the range $1\leq
r<\alpha^{-1}$.  

By expressing the scattering parameters for the full device (defined by ports
$1$, $2$, $3$, $4$) in terms of the scattering parameters for the inner device
(defined by ports $1^{\prime}$, $2^{\prime}$, $3$, $4$), we get
\begin{align}
\left[  S\right]    & =\left(
\begin{array}
[c]{cccc}%
S_{11} & S_{12} & S_{13} & S_{14}\\
S_{21} & S_{22} & S_{23} & S_{24}\\
S_{31} & S_{32} & S_{33} & S_{34}\\
S_{41} & S_{42} & S_{43} & S_{44}%
\end{array}
\right)  \nonumber\\
& =\left(
\begin{array}
[c]{cccc}%
\frac{1}{2}\left(  s_{1^{\prime}1^{\prime}}-s_{2^{\prime}2^{\prime}%
}+is_{2^{\prime}1^{\prime}}+is_{1^{\prime}2^{\prime}}\right)   & \frac{1}%
{2}\left(  is_{1^{\prime}1^{\prime}}+is_{2^{\prime}2^{\prime}}+s_{1^{\prime
}2^{\prime}}-s_{2^{\prime}1^{\prime}}\right)   & \frac{1}{\sqrt{2}}\left(
is_{2^{\prime}3}+s_{1^{\prime}3}\right)   & \frac{1}{\sqrt{2}}\left(
is_{2^{\prime}4}+s_{1^{\prime}4}\right)  \\
\frac{1}{2}\left(  is_{1^{\prime}1^{\prime}}+is_{2^{\prime}2^{\prime}%
}+s_{2^{\prime}1^{\prime}}-s_{1^{\prime}2^{\prime}}\right)   & \frac{1}%
{2}\left(  s_{2^{\prime}2^{\prime}}-s_{1^{\prime}1^{\prime}}+is_{2^{\prime
}1^{\prime}}+is_{1^{\prime}2^{\prime}}\right)   & \frac{1}{\sqrt{2}}\left(
s_{2^{\prime}3}+is_{1^{\prime}3}\right)   & \frac{1}{\sqrt{2}}\left(
s_{2^{\prime}4}+is_{1^{\prime}4}\right)  \\
\frac{1}{\sqrt{2}}\left(  is_{32^{\prime}}+s_{31^{\prime}}\right)   & \frac
{1}{\sqrt{2}}\left(  s_{32^{\prime}}+is_{31^{\prime}}\right)   & s_{33} &
s_{34}\\
\frac{1}{\sqrt{2}}\left(  is_{42^{\prime}}+s_{41^{\prime}}\right)   & \frac
{1}{\sqrt{2}}\left(  s_{42^{\prime}}+is_{41^{\prime}}\right)   & s_{43} &
s_{44}%
\end{array}
\right),  \label{S_mat}%
\end{align}

where the scattering parameters of the inner device (without the front
coupler) are given by%

\begin{align}
\left[  s\right]    & =\left(
\begin{array}
[c]{cccc}%
s_{1^{\prime}1^{\prime}} & s_{1^{\prime}2^{\prime}} & s_{1^{\prime}3} &
s_{1^{\prime}4}\\
s_{2^{\prime}1^{\prime}} & s_{2^{\prime}2^{\prime}} & s_{2^{\prime}3} &
s_{2^{\prime}4}\\
s_{31^{\prime}} & s_{32^{\prime}} & s_{33} & s_{34}\\
s_{41^{\prime}} & s_{42^{\prime}} & s_{43} & s_{44}%
\end{array}
\right)  \nonumber\\
& =\left(
\begin{array}
[c]{cccc}%
\frac{r\beta^{2}}{1-\alpha^{2}r^{2}} & \frac{\alpha s^{2}e^{-i\varphi_{12}}%
}{1-\alpha^{2}r^{2}} & \frac{i\beta se^{-i\varphi_{1}}}{1-\alpha^{2}r^{2}} &
\frac{i\beta r\alpha se^{-i\varphi_{1}}}{1-\alpha^{2}r^{2}}\\
\frac{\alpha s^{2}e^{i\varphi_{12}}}{1-\alpha^{2}r^{2}} & \frac{r\beta^{2}%
}{1-\alpha^{2}r^{2}} & \frac{i\beta r\alpha se^{-i\varphi_{2}}}{1-\alpha
^{2}r^{2}} & \frac{i\beta se^{-i\varphi_{2}}}{1-\alpha^{2}r^{2}}\\
\frac{i\beta se^{i\varphi_{1}}}{1-\alpha^{2}r^{2}} & \frac{i\beta r\alpha
se^{i\varphi_{2}}}{1-\alpha^{2}r^{2}} & -\frac{\beta^{2}r}{1-\alpha^{2}r^{2}}
& \frac{\alpha s^{2}}{1-\alpha^{2}r^{2}}\\
\frac{i\beta r\alpha se^{i\varphi_{1}}}{1-\alpha^{2}r^{2}} & \frac{i\beta
se^{i\varphi_{2}}}{1-\alpha^{2}r^{2}} & \frac{\alpha s^{2}}{1-\alpha^{2}r^{2}}
& -\frac{\beta^{2}r}{1-\alpha^{2}r^{2}}%
\end{array}
\right)  .\label{s_mat}%
\end{align}

It is worthwhile noting that the scattering matrix $[s]$ (Eq. (\ref{s_mat})) is symplectic (information preserving). This can be seen  for instance from 

\begin{equation}
\left\vert s_{1^{\prime}1^{\prime}}\right\vert ^{2}+\left\vert s_{1^{\prime
}2^{\prime}}\right\vert ^{2}-\left\vert s_{1^{\prime}3}\right\vert
^{2}-\left\vert s_{1^{\prime}4}\right\vert ^{2}=1.
\end{equation}

Moreover, since $[s]$ is symplectic and the front coupler is a unitary element, it follows that the scattering matrix of the whole device Eq. (\ref{S_mat}) is symplectic as well.

By substituting the scattering parameters given in Eq. (\ref{s_mat}) into Eq. (\ref{S_mat}) and writing the expressions in terms of the parameter $s$, we get the
scattering parameters of the JDA at resonance in an explicit form %

\begin{align}
S_{21} &  =\frac{i}{1-\frac{\alpha^{2}}{\beta^{2}}s^{2}}\left[  \sqrt{1+s^{2}%
}+\frac{\alpha}{\beta^{2}}s^{2}\sin\varphi_{12}\right]  ,\label{S21}\\
S_{12} &  =\frac{i}{1-\frac{\alpha^{2}}{\beta^{2}}s^{2}}\left[  \sqrt{1+s^{2}%
}-\frac{\alpha}{\beta^{2}}s^{2}\sin\varphi_{12}\right]  ,\label{S12}\\
S_{11} &  =S_{22}=\frac{i\alpha}{\beta^{2}}\frac{s^{2}}{1-\frac{\alpha^{2}%
}{\beta^{2}}s^{2}}\cos\varphi_{12},\label{S11}\\
S_{33} &  =S_{44}=-\frac{\sqrt{1+s^{2}}}{1-\frac{\alpha^{2}}{\beta^{2}}s^{2}%
},\label{S33}\\
S_{34} &  =S_{43}=\frac{\alpha}{\beta^{2}}\frac{s^{2}}{1-\frac{\alpha^{2}%
}{\beta^{2}}s^{2}},\label{S34}\\
S_{31} &  =\frac{se^{i\left(  \varphi_{1}+\varphi_{2}\right)  /2+3i\pi/4}%
}{\sqrt{2}\beta\left(  1-\frac{\alpha^{2}}{\beta^{2}}s^{2}\right)  }\left[
\sqrt{1+s^{2}}\alpha e^{-i\frac{\varphi_{12}}{2}+i\frac{\pi}{4}}%
+e^{i\frac{\varphi_{12}}{2}-i\frac{\pi}{4}}\right]  ,\label{S31}\\
S_{41} &  =\frac{se^{i\left(  \varphi_{1}+\varphi_{2}\right)  /2+3i\pi/4}%
}{\sqrt{2}\beta\left(  1-\frac{\alpha^{2}}{\beta^{2}}s^{2}\right)  }\left[
e^{-i\frac{\varphi_{12}}{2}+i\frac{\pi}{4}}+\sqrt{1+s^{2}}\alpha
e^{i\frac{\varphi_{12}}{2}-i\frac{\pi}{4}}\right]  ,\label{S41}\\
S_{32} &  =\frac{se^{i\left(  \varphi_{1}+\varphi_{2}\right)  /2+i3\pi/4}%
}{\sqrt{2}\beta\left(  1-\frac{\alpha^{2}}{\beta^{2}}s^{2}\right)  }\left[
\sqrt{1+s^{2}}\alpha e^{-i\frac{\varphi_{12}}{2}-i\frac{\pi}{4}}%
+e^{i\frac{\varphi_{12}}{2}+i\frac{\pi}{4}}\right]  ,\label{S32}\\
S_{42} &  =\frac{se^{i\left(  \varphi_{1}+\varphi_{2}\right)  /2+i3\pi/4}%
}{\sqrt{2}\beta\left(  1-\frac{\alpha^{2}}{\beta^{2}}s^{2}\right)  }\left[
e^{-i\frac{\varphi_{12}}{2}-i\frac{\pi}{4}}+\sqrt{1+s^{2}}\alpha
e^{i\frac{\varphi_{12}}{2}+i\frac{\pi}{4}}\right]  ,\label{S42}\\
S_{14} &  =\frac{se^{-i\left(  \varphi_{2}+\varphi_{1}\right)  /2+i3\pi/4}%
}{\sqrt{2}\beta\left(  1-\frac{\alpha^{2}}{\beta^{2}}s^{2}\right)  }\left[
e^{i\frac{\varphi_{12}}{2}+i\frac{\pi}{4}}+\sqrt{1+s^{2}}\alpha e^{-i\frac
{\varphi_{12}}{2}-i\frac{\pi}{4}}\right]  ,\label{S14}\\
S_{13} &  =\frac{se^{-i\left(  \varphi_{2}+\varphi_{1}\right)  /2+i3\pi/4}%
}{\sqrt{2}\beta\left(  1-\frac{\alpha^{2}}{\beta^{2}}s^{2}\right)  }\left[
\sqrt{1+s^{2}}\alpha e^{i\frac{\varphi_{12}}{2}+i\frac{\pi}{4}}+e^{-i\frac
{\varphi_{12}}{2}-i\frac{\pi}{4}}\right]  ,\label{S13}\\
S_{23} &  =\frac{se^{-i\left(  \varphi_{2}+\varphi_{1}\right)  /2+i3\pi/4}%
}{\sqrt{2}\beta\left(  1-\frac{\alpha^{2}}{\beta^{2}}s^{2}\right)  }\left[
\sqrt{1+s^{2}}\alpha e^{i\frac{\varphi_{12}}{2}-i\frac{\pi}{4}}+e^{-i\frac
{\varphi_{12}}{2}+i\frac{\pi}{4}}\right]  ,\label{S23}\\
S_{24} &  =\frac{se^{-i\left(  \varphi_{2}+\varphi_{1}\right)  /2+i3\pi/4}%
}{\sqrt{2}\beta\left(  1-\frac{\alpha^{2}}{\beta^{2}}s^{2}\right)  }\left[
e^{i\frac{\varphi_{12}}{2}-i\frac{\pi}{4}}+\sqrt{1+s^{2}}\alpha e^{-i\frac
{\varphi_{12}}{2}+i\frac{\pi}{4}}\right]  ,\label{S24}%
\end{align}

where $\varphi_{12}\equiv\varphi_{1}-\varphi_{2}.$

\begin{figure}
[h]
\begin{center}
\includegraphics[
width=0.55\textwidth
]%
{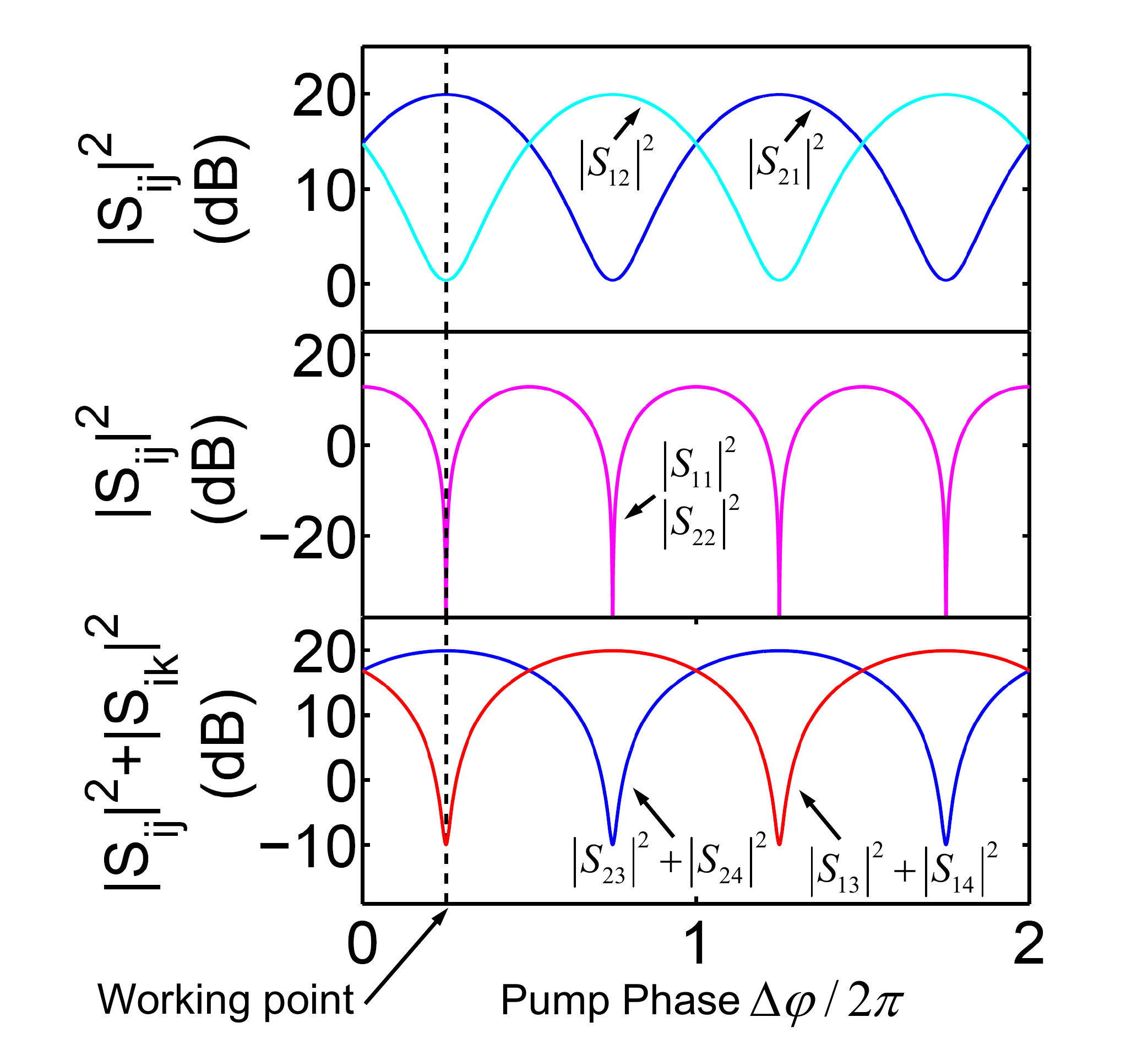}%
\caption{(color online). Calculated scattering parameters of the JDA, shown in Fig. \ref{DAScheme}, as a function of the pump phase difference. The calculation was performed at resonance for the case of a 3 dB back coupler (i.e. $\alpha=\beta=1/\sqrt{2}$) and r=1.326 (which yields a gain of 20 dB in the forward direction). The top panel displays the forward ($|S_{21}|^2$) and backward gains ($|S_{12}|^2$) of the amplifier using blue and cyan curves respectively. The middle panel displays both reflection gains (i.e. $|S_{11}|^2$ and $|S_{22}|^2$) using a magenta curve, while the bottom panel exhibits the effect of the cold loads (ports 3 and 4) on the input (1) and output (2) ports. The vertical dashed black line corresponding to $\varphi_{12}=\pi/2$ indicates a desired working point of the JDA. To generate the different curves equations (\ref{S21}), (\ref{S12}), (\ref{S11}), (\ref{S14}), (\ref{S13}), (\ref{S23}), (\ref{S24}) have been employed.} %
\label{ThySparamPumpPhase}%
\end{center}
\end{figure}

In Fig. \ref{ThySparamPumpPhase} we display a calculation result for selected scattering parameters of the JDA as a function of the pump phase difference. The calculation assumes a 3 dB back coupler ($\alpha=\beta=1/\sqrt{2}$). The top panel shows the gain of the device in the forward $|S_{21}|^2$ ($1\rightarrow2$) and reverse $|S_{12}|^2$ ($2\rightarrow1$) direction. The middle panel shows the reflection parameters off ports $1$ and $2$ ($|S_{11}|^2$ and $|S_{22}|^2$). While the bottom panel exhibits the effect of the dumps, i.e. ports $3$ and $4$, on the input and output ports. The blue and red curves depict the sum $|S_{23}|^2+|S_{24}|^2$ and $|S_{13}|^2+|S_{14}|^2$ respectively. In this calculation, the reflection-gain amplitude $r$ of the JDA was chosen to give a maximum gain of 20 dB in the forward direction. As can be seen in the figure, the gain in this direction is maximized for $\varphi_{12}=\pi/2$. Also, at this working point indicated by the vertical dashed line, $|S_{12}|^2$ is minimal ($0.42$ dB), the reflection parameters $|S_{11}|^2$ and $|S_{22}|^2$ vanish, and signals originating from the dumps are attenuated. The fact that the sum $|S_{23}|^2+|S_{24}|^2$ at this working point has an identical gain as $|S_{21}|^2$ further shows that the JDA is quantum-limited with a minimum amount of added noise coming from the dumps \cite{Caves}.

It is straightforward to verify that for a pump phase difference of $\varphi_{12}=\pi/2$, the scattering matrix reduces into%

\begin{equation}
\left[  S\right]  =\left(
\begin{array}
[c]{cccc}%
0 & i\left(  g-h\right)   & \sqrt{\frac{\left(  g-h\right)  ^{2}-1}{2}} &
-\sqrt{\frac{\left(  g-h\right)  ^{2}-1}{2}}\\
i\left(  g+h\right)   & 0 & i\sqrt{\frac{\left(  g+h\right)  ^{2}-1}{2}} &
i\sqrt{\frac{\left(  g+h\right)  ^{2}-1}{2}}\\
-\sqrt{\frac{\left(  g+h\right)  ^{2}-1}{2}} & -i\sqrt{\frac{\left(
g-h\right)  ^{2}-1}{2}} & -g & h\\
-\sqrt{\frac{\left(  g+h\right)  ^{2}-1}{2}} & i\sqrt{\frac{\left(
g-h\right)  ^{2}-1}{2}} & h & -g
\end{array}
\right)  ,\label{S_mat_symm_back_coup}%
\end{equation}

where 

\begin{align}
g &  =\frac{\sqrt{1+s^{2}}}{1-\frac{\alpha^{2}}{\beta^{2}}s^{2}},\label{gWalpha}\\
h &  =\frac{\alpha}{\beta^{2}}\frac{s^{2}}{1-\frac{\alpha^{2}}{\beta^{2}}s^{2}}.\label{hWalpha}%
\end{align}

In the special case where the back coupler is balanced, i.e. $\alpha=\beta=1/\sqrt{2}$, we get

\begin{align}
g &  =\frac{\sqrt{1+s^{2}}}{1-s^{2}},\label{g}\\
h &  =\frac{\sqrt{2}s^{2}}{1-s^{2}}.\label{h}%
\end{align}

\begin{figure}
[h]
\begin{center}
\includegraphics[
width=0.55\textwidth
]%
{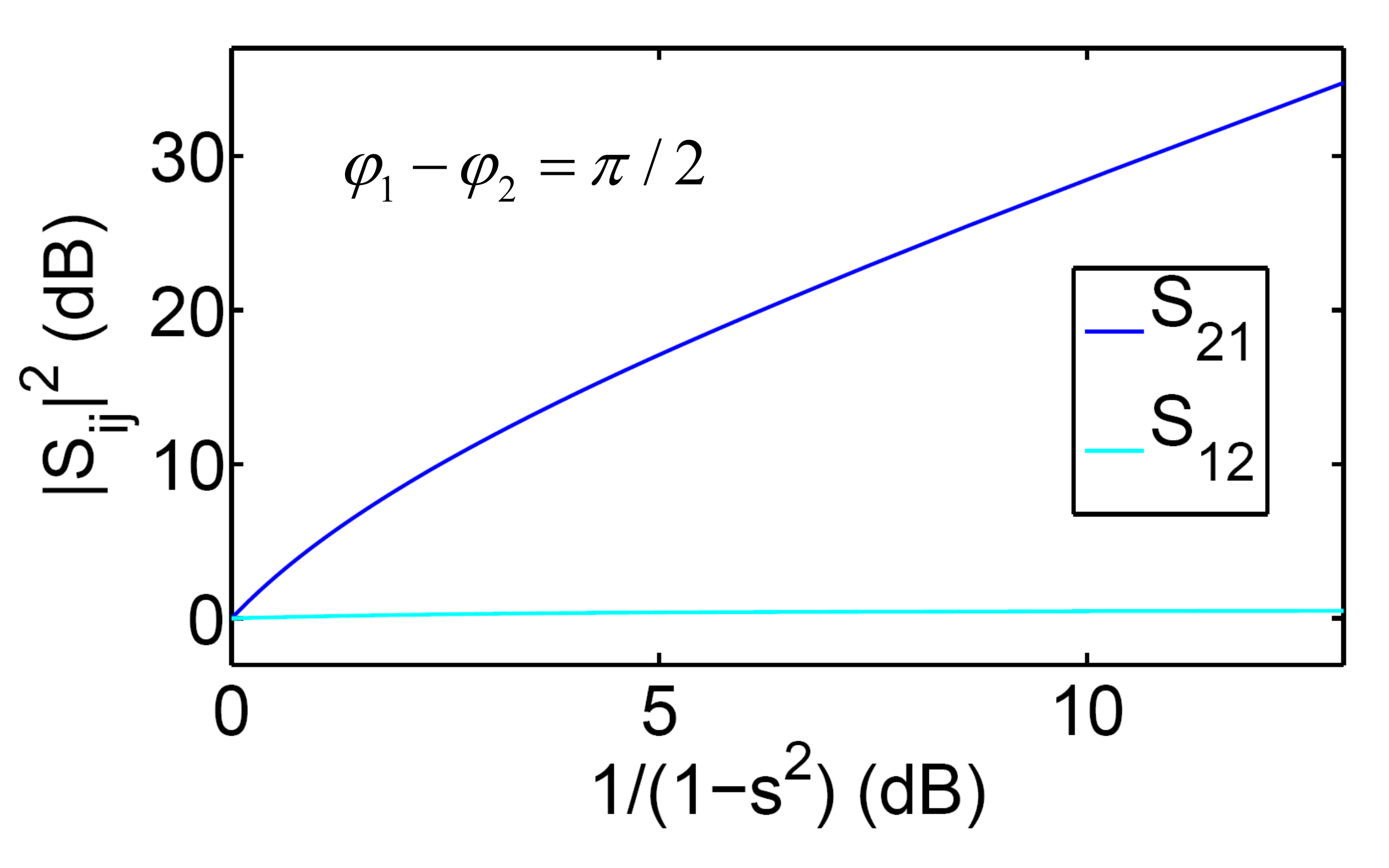}%
\caption{(color online). Calculated dependence of $|S_{21}|^2$ (blue) and $|S_{12}|^2$ (cyan) on the trans-gain amplitude $s$ at resonance (expressed as $1/(1-s^2)$). The calculation was carried out for the case of an ideal JDA shown in Fig. \ref{DAScheme} at a pump phase difference $\varphi_{12}=\pi/2$. The calculation also assumes a $3$ dB back coupler  (i.e. $\alpha=\beta=1/\sqrt{2}$).}%
\label{ThyS21S12vsS}%
\end{center}
\end{figure}

It is worthwhile noting that in the limit of $s\rightarrow1^{-}$ (high gain limit) we get 

\begin{align}
\lim_{s\rightarrow1^{-}}g+h  & =+\infty,\label{gPhsTo1}\\
\lim_{s\rightarrow1^{-}}g-h  & =\frac{3}{2\sqrt{2}},\label{gMhsTo1}%
\end{align}

where Eq. (\ref{gMhsTo1}) sets a bound on the reverse gain of the device $|S_{12}|^2\leq9/8$. For $s=0$ (amplifier off) we obtain 

\begin{align}
g+h  & =1,\label{gPhsTo0}\\
g-h  & =1, \label{gPhsTo0}
\end{align}

and the scattering matrix of the device becomes

\begin{equation}
\left[  S\right]  =\left(
\begin{array}
[c]{cccc}%
0 & i & 0 & 0\\
i & 0 & 0 & 0\\
0 & 0 & -1 & 0\\
0 & 0 & 0 & -1
\end{array}
\right) \cdot\label{S_mat_s_zero}
\end{equation}

This result is particularly important in the experiment as it shows that when the JDA is off (no pumps are applied) signals can propagate between ports 1 and 2 with unity transmission (accompanied with a certain phase shift). Hence, the qubit connected to port 1 of the JDA can be measured directly without disconnecting the JDA or using a switch.
 
Furthermore, by defining the relations%

\begin{align}
g+h &  =\sqrt{G},\label{sqrtG}\\
g-h &  =\sqrt{H},\label{sqrtH}%
\end{align}

we can cast Eq. (\ref{S_mat_symm_back_coup}) in the simple form%

\begin{equation}
\left[  S\right]  =\left(
\begin{array}
[c]{cccc}%
0 & i\sqrt{H} & \sqrt{\frac{H-1}{2}} & -\sqrt{\frac{H-1}{2}}\\
i\sqrt{G} & 0 & i\sqrt{\frac{G-1}{2}} & i\sqrt{\frac{G-1}{2}}\\
-\sqrt{\frac{G-1}{2}} & -i\sqrt{\frac{H-1}{2}} & -\frac{\sqrt{G}+\sqrt{H}}{2}
& \frac{\sqrt{G}-\sqrt{H}}{2}\\
-\sqrt{\frac{G-1}{2}} & i\sqrt{\frac{H-1}{2}} & \frac{\sqrt{G}-\sqrt{H}}{2} &
-\frac{\sqrt{G}+\sqrt{H}}{2}%
\end{array}
\right)  \label{S_mat_G_H}%
\end{equation}

displaying symplecticity in an obvious manner.

\begin{figure}
[h]
\begin{center}
\includegraphics[
width=0.55\textwidth
]%
{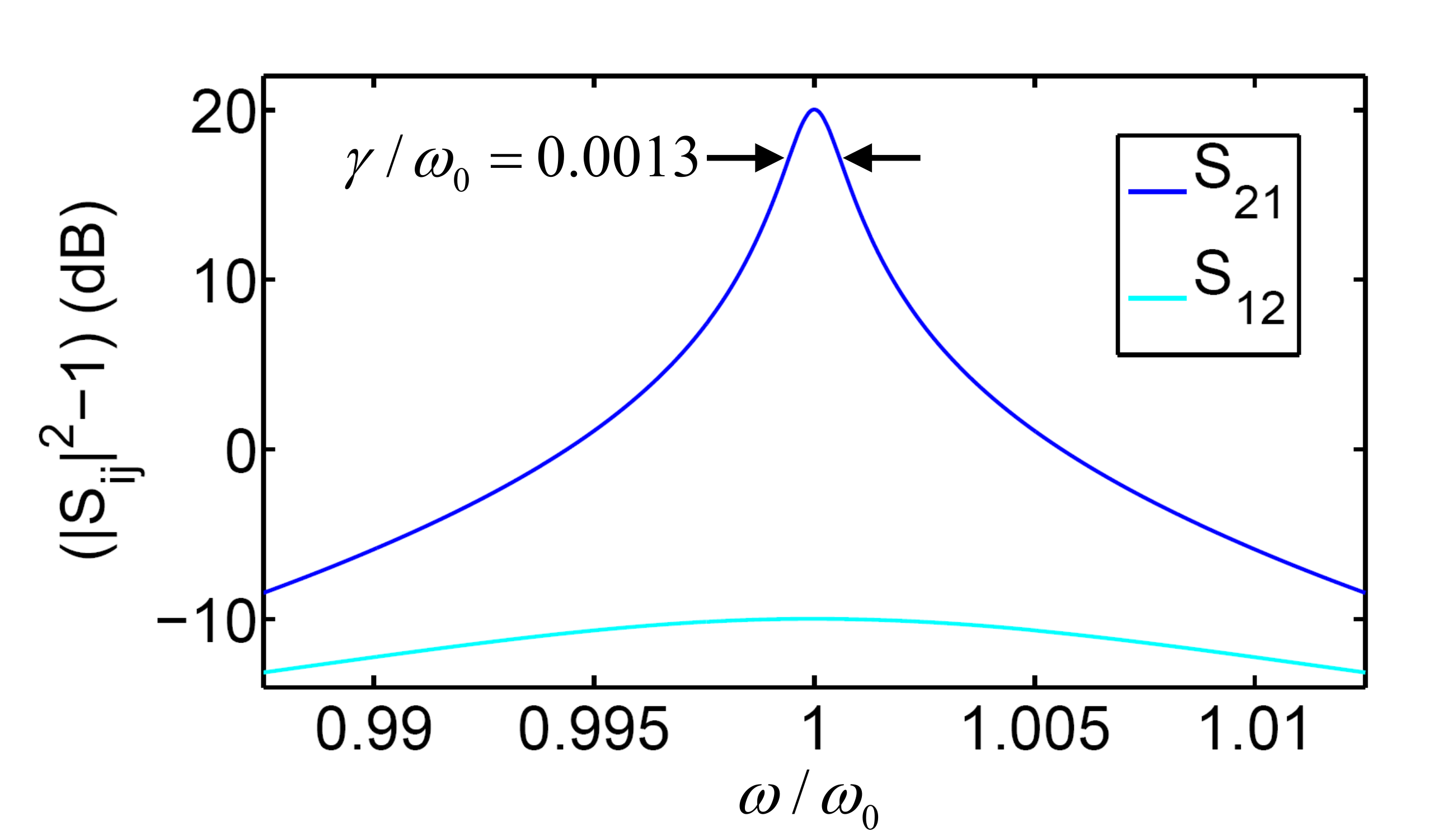}%
\caption{(color online). Calculated $S_{21}$ (blue) and $S_{12}$ (cyan) response versus normalized frequency for an ideal JDA shown in Fig. \ref{DAScheme}. The calculation is carried out for the case of a 3 dB back coupler (i.e. $\alpha=\beta=1/\sqrt{2}$) and for a pump phase difference $\varphi_{12}=\pi/2$. The notations $\omega_{0}$ and $\gamma$ correspond to the angular resonance frequency of the signal resonator and the dynamical bandwidth (3 dB point from the peak) of the amplifier. Other parameters employed in the calculation are $r=1.326$ (which yields a gain of $20$ dB at resonance), the angular resonance frequency of the idler resonator $1.875\omega_{0}$, the angular frequency of the pump drive $2.875\omega_{0}$, and the bandwidths of the signal and idler resonators $0.025\omega_{0}$ and $0.05\omega_{0}$ respectively.}%
\label{ThyFreqResp}%
\end{center}
\end{figure}

\begin{figure}
[h]
\begin{center}
\includegraphics[
width=0.55\textwidth
]%
{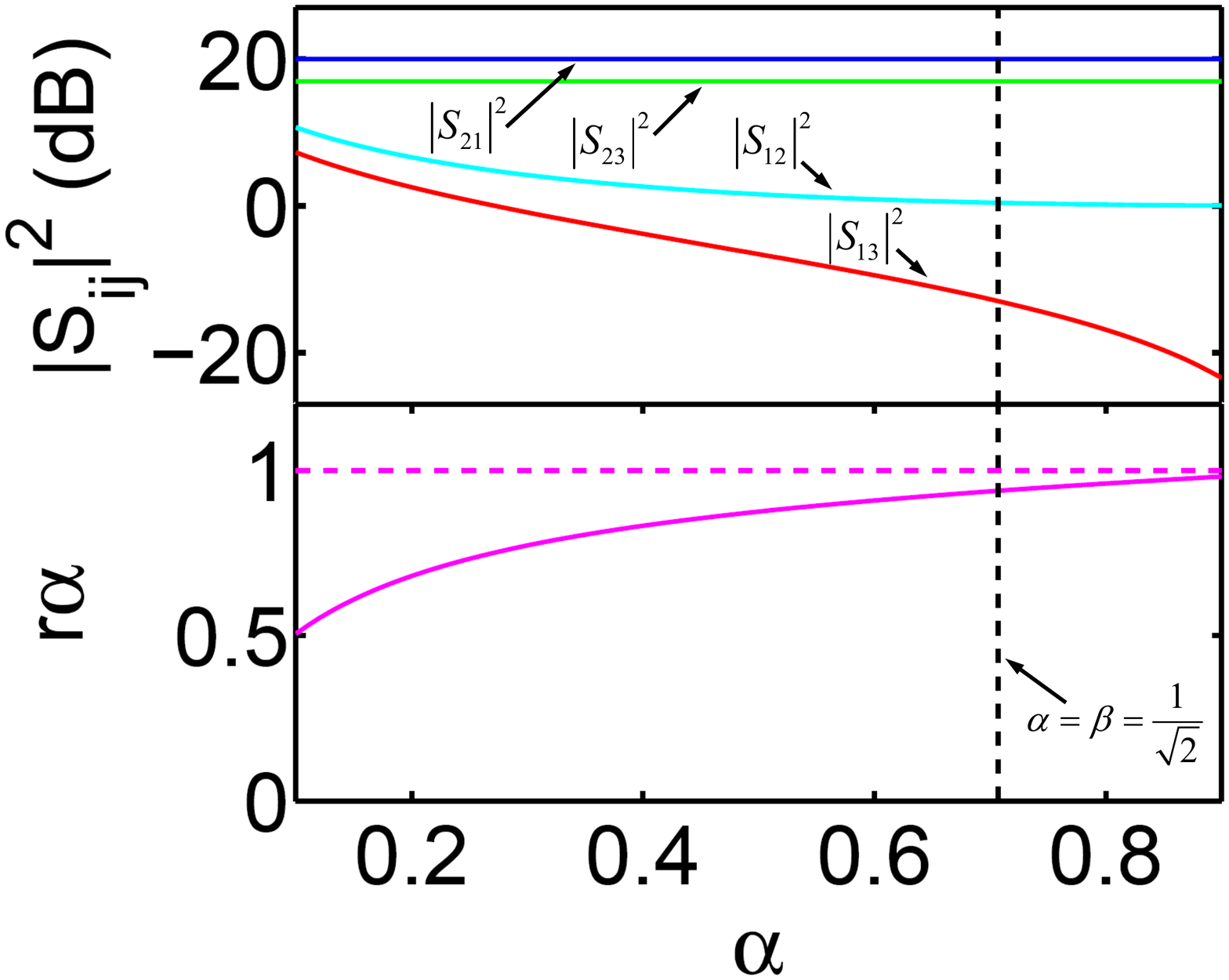}%
\caption{(color online). Calculated scattering parameters of the JDA (depicted in Fig. \ref{DAScheme}) as a function of the coefficient $\alpha$ of the back coupler. In this calculation, the reflection amplitude of the paramps was set to yield a constant forward gain ($|S_{21}|^2$) of 20 dB for each value of $\alpha$. The values of the parameter $r$ can be inferred from the bottom panel where the product $r\alpha$ is drawn. The horizontal dashed line indicates an onset of instability of the device (where $r\alpha=1$). The curves $|S_{12}|^2$, $|S_{23}|^2$, and $|S_{13}|^2$ drawn in the top panel can be used in order to quantify different figures of merit for the system, i.e. reverse gain, added noise, and amplified noise at the input due to the dumps. The vertical dashed line in both panels corresponds to the case of a 3 dB back coupler.}%
\label{SparamVsAlpha}%
\end{center}
\end{figure}
 
In Fig. \ref{ThyS21S12vsS}, we plot the dependence of the forward and backward gains of the JDA at resonance at $\varphi_{12}=\pi/2$ as a function of the parameter $s$ (drawn versus $1/(1-s^2)$). The calculation is performed for a Josephson directional amplifier with a $3$ dB back coupler. As can be seen in the figure, while the forward gain $|S_{21}|^2$ diverges in the limit of $s\rightarrow1^{-}$, $|S_{12}|^2$ reaches a plateau ($0.5$ dB). Furthermore, in Fig. \ref{ThyFreqResp} we plot the calculated response of the forward and backward gains of the JDA at the desired working point (i.e. $\varphi_{12}=\pi/2$) as a function of normalized frequency $\omega/\omega_{0}$, where $\omega_{0}$ is the angular resonance frequency of the signal resonator of the paramp. Similar to Figs. \ref{ThySparamPumpPhase} and \ref{ThyS21S12vsS} this calculation is performed for a $3$ dB back coupler as well. In order to obtain the frequency response of the device, we use the dependence of the reflection and trans-gain amplitudes of the paramp (i.e. Josephson parametric converter) on frequency (see full equations in Ref. \cite{JRCAreview}) and solve the flow graph of Fig. \ref{DAScheme} for each excitation frequency $\omega$ using the calculation method outlined in Ref. \cite{BDA}. The frequency dependence calculation also assumes for simplicity that the response of the front and back couplers is frequency independent (in practice this assumption implies that the bandwidth of the couplers is very large compared to the bandwidth of the Josephson parametric converters used in the scheme). 

Moreover, in order to quantify the dependence of the JDA performance on the splitting parameters of the back coupler $\alpha$ and $\beta$, we plot in Fig. \ref{SparamVsAlpha} (top panel) graphs of selected scattering parameters of the device as a function of $\alpha$. The scattering parameters are calculated at resonance and $\varphi_{12}=\pi/2$. For each value of $\alpha$ (varied in the range $0.1$ and $0.9$) we calculate the corresponding reflection amplitude $r$ which gives a forward gain ($|S_{21}|^2$) of $20$ dB. The corresponding values of $r$ can be inferred from the graph of $r\alpha$ plotted in the bottom panel. As can be seen in the top panel, for a fixed $|S_{21}|^2$ of $20$ dB, decreasing $\alpha$ (i.e. increasing the attenuation in the JDA gain loop) yields a slow increase in the reverse gain $|S_{12}|^2$ and in the effect of incoming signals from the dumps (i.e. noise) on the input (characterized by $|S_{13}|^2$, since $|S_{14}|^2$=$|S_{13}|^2$). Also, the fact that $|S_{23}|^2$ is constant and lies $3$ dB lower than $|S_{21}|^2$ (note that $|S_{24}|^2$=$|S_{23}|^2$) verifies that the JDA is quantum-limited independently on the value of $\alpha$. This is because the added noise by the JDA (which originates from quantum noise at the dumps) gets amplified at the output by the same amount as quantum noise at the input. The vertical dashed line outlines the special case of a $3$ dB back coupler (used in Figs. \ref{ThySparamPumpPhase}, \ref{ThyS21S12vsS}, and \ref{ThyFreqResp}), where $\alpha=\beta=1/\sqrt{2}$. From this figure we can see that although it is advantageous to choose $\alpha$ very close to unity in order to keep $|S_{12}|^2$ and $|S_{13}|^2$ as small as possible for a fixed gain $|S_{21}|^2$, however since the product $r\alpha$ approaches unity as well, the JDA will be more susceptible to instability. Hence, the case of a $3$ dB back coupler is shown since it offers a convenient compromise between the minimization of $S_{12}$ and gain stability.

\section{JDA circuit} 

\begin{figure}
[htb]
\begin{center}
\includegraphics[
width=0.55\textwidth 
]
{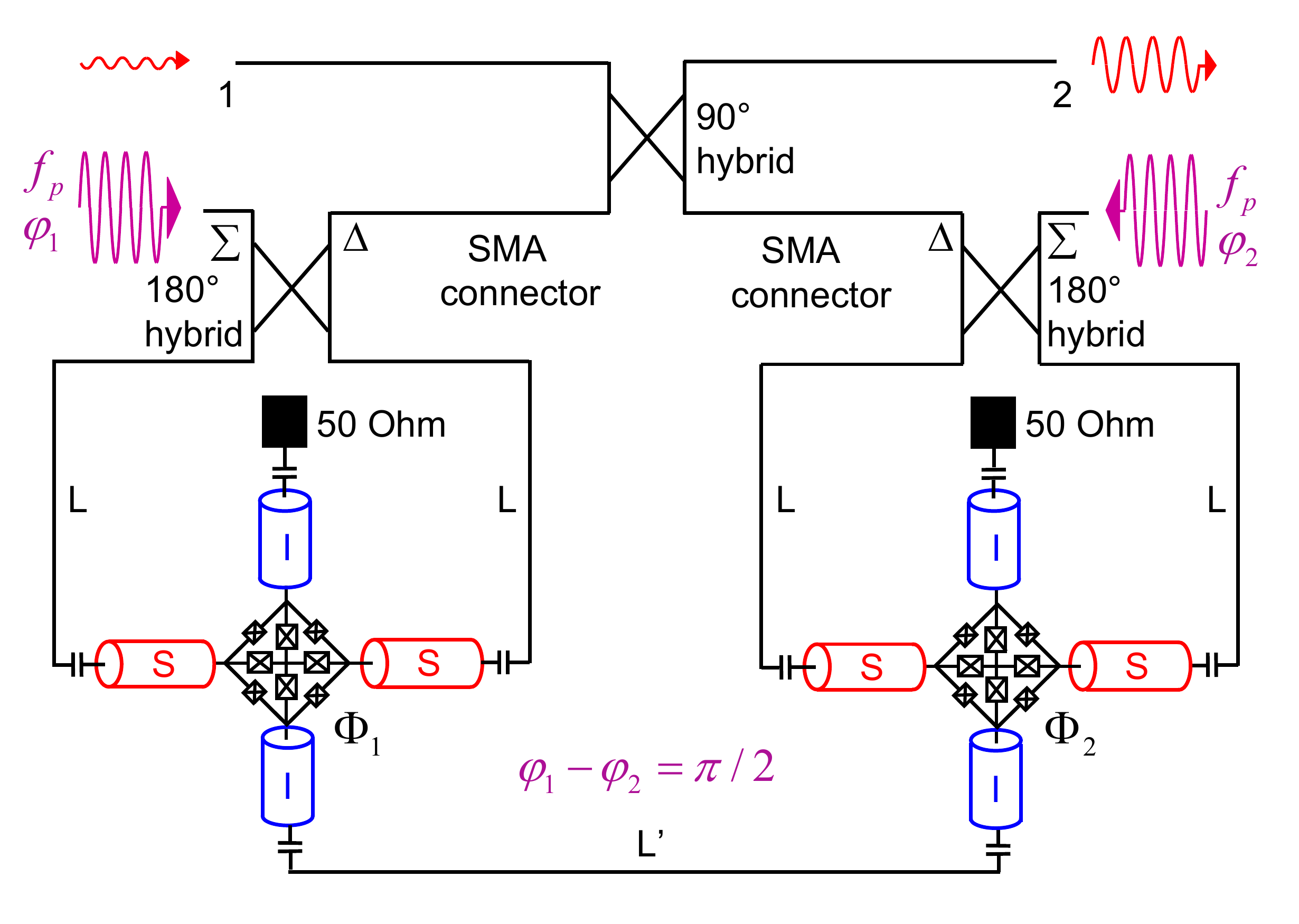}
\caption{(color online). A diagram of the JDA circuit used in the experiment.}
\label{DAcircuit}
\end{center}
\end{figure}

In Fig. \ref{DAcircuit} we show a diagram of the JDA circuit that has been used in the experiment (see main text). The circuit consists of two nominally identical microstrip JPCs \cite{Jamp} that are coupled together through their signal (S) and idler (I) ports. The signal feedlines of each JPC are connected to a commercial 180 degree hybrid using phase-matched copper coax cables of length L=$2"$, with a phase mismatch of less than $1.5$ degrees at the signal tone frequency. The signal ports of the two JPCs, defined at the plane of the difference port ($\Delta$) of the 180 degree hybrids, are connected together using a commercial 90 degree hybrid which in turn defines the JDA ports $1$ and $2$. On the side of the idler feedlines, we substituted the back coupler shown in the generic scheme in Fig. \ref{DAScheme} with a semi-equivalent circuit for practical reasons (in order to decrease the number of components and save in volume). We connected one feedline of each idler resonator to a 50 Ohm cold termination, and connected the two remaining ones with a copper coax cable of length L'=$6"$ (which has an insertion loss of about $0.52$ dB measured at the idler tone frequency at $10.75$ $\operatorname{GHz}$). Since the idler resonators are coupled to the feedlines through equal coupling capacitors on each side, connecting one of them to $50$ Ohm termination, results in a $3$ dB power loss for the outgoing waves on the idler port (similar to the effect generated by a $3$ dB back coupler). The pumps which are strong non-resonant common drives at frequency $f_{p}$ are fed to the JPCs through the sum ports of the 180 degree hybrids. In order to achieve directionality, a phase difference of $\pi/2$ is applied between the two coherent pump drives feeding the device.

The Josephson ring modulator (JRM), which is positioned at the center of each JPC at an rf-current anti-node, consists of 8 Josephson junctions. The inner 4 junctions which act as linear inductors are added to allow frequency tunability of each JPC by threading external flux into the loop \cite{Roch,JRCAreview}. The external flux is applied to the JRM using a small superconducting magnetic coil that is attached to the copper box housing the JPC. Using this configuration, we are able to tune the center frequency of the JDA over more than $150$ $\operatorname{MHz}$.   

The two JPCs have been designed with the same parameters, fabricated on the same silicon wafer, and diced into two separate chips at the end of the fabrication process. Each chip was mounted and housed in a different copper box. The experiment was carried out in $3$ consecutive cooldowns. In the first cooldown, the two JPCs were measured and characterized individually. In the second cooldown, they were combined together in a JDA configuration as shown in Fig. \ref{DAcircuit}, and measured using two circulators connected to both ports. This setup allowed us to probe all four scattering parameters of the device and characterize it with respect to gain, directionality, matching to the input and output, bandwidth, SNR improvement, dynamic range, and tunability (mentioned earlier). In this cooldown, we were able to verify all the important features of the JDA predicted by theory. As a particular example, we obtained at a center frequency of $7.695$ $\operatorname{GHz}$ and a pump phase difference of $\pi/2$, a forward gain of about $21$ dB ($|S_{21}|^2$), a reverse gain of less than $1$ dB ($|S_{12}|^2$), reflection attenuation of about $-10$ dB in $|S_{11}|^2$ and $|S_{22}|^2$, a dynamical bandwidth of $6$ $\operatorname{MHz}$, SNR improvement of $16$ dB in the forward direction, and a maximum input power of about $1$ photon at the signal frequency per inverse dynamical bandwidth. In the third cooldown, we coupled the JDA to a previously measured cavity-qubit system as shown in Fig. 2 in the main text. In that setup, we lost the option of measuring $S_{11}$ and $S_{12}$ directly, which limited our ability to tune the JDA to some extent. Also, with that setup, we were not able to operate the JDA above $12$ dB of forward gain in a stable manner. This can be due to, (1) a change in the impedance seen by port 1 of the JDA when it is connected directly to the cavity-qubit system through a directional coupler and a low-pass filter compared to the $50$ Ohm port of a circulator, and (2) a change in the impedance of the pump lines (which causes reflections) as a result of adding infra red filters to all input and output lines of the setup including the pump lines, in order to protect the cavity-qubit system.         

Possible improvements to the present JDA configuration include (1) substituting the commercial hybrids with superconducting on-chip versions \cite{LehnertCoupler} which have less insertion loss, less phase and amplitude imbalance between the different arms, and larger isolation between the sum and difference ports in the case of the 180 degree hybrids, (2) substituting the normal coax cable connecting between the two idler feedlines with a superconducting transmission line, and of course ultimately (3) implementing all components on the same chip. 

\section{JDA working point}

\begin{figure}
[h]
\begin{center}
\includegraphics[
width=0.55\textwidth
]%
{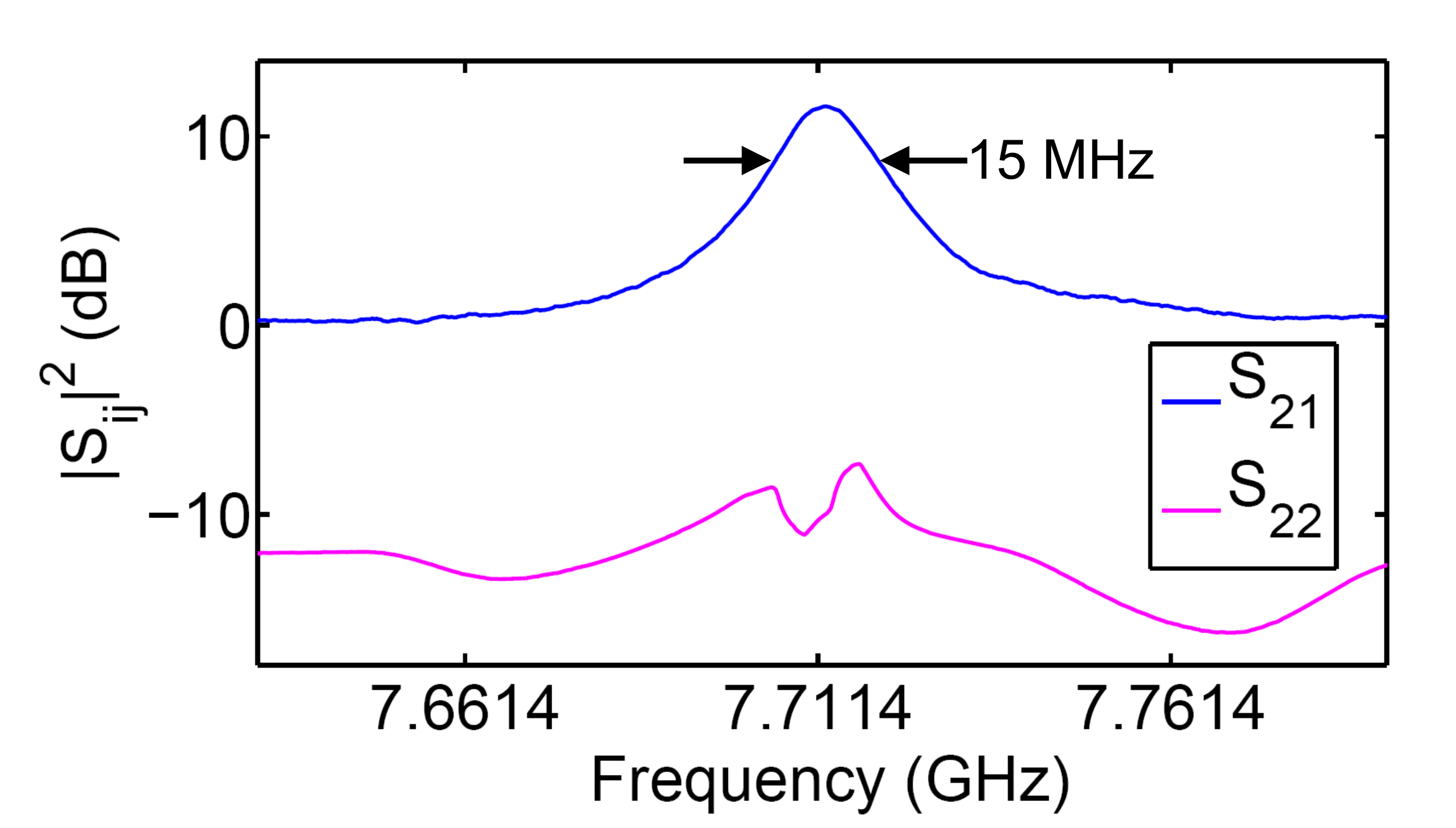}%
\caption{(color online). Measured $S_{21}$ (blue) and $S_{22}$ (magenta) of the JDA as a function of frequency. The measurement was taken with a vector network analyzer at the working point used in the readout of the qubit state (as described in the main text). The input lines Probe$_{1}$ and Probe$_{2}$ indicated in Fig. 2 in the main text were used in the measurement of $S_{21}$ and $S_{22}$ respectively.}%
\label{GainExp}%
\end{center}
\end{figure}

In Fig. \ref{GainExp} we show vector network analyzer (VNA) measurements of the scattering parameters $S_{21}$ (blue) and $S_{22}$ (magenta) of the JDA as a function of frequency at the working point employed in the readout of the qubit-cavity system (see main text). To measure these parameters, a weak coherent signal was applied through port 1 of the VNA to Probe$_{1}$ and Probe$_{2}$ input lines of the experimental setup shown in Fig. 2 (see main text). Port 2 of the VNA was connected to the output line after bypassing the demodulation scheme used in the qubit measurements (located outside of the fridge). At this working point, the measured gain in the forward direction ($|S_{21}|^2$) and the dynamical bandwidth (defined at the 3 dB points below the maximum) are $11.6\pm1$ dB  and $15$ $\operatorname{MHz}$ respectively.